\documentclass[aps,prl,twocolumn,amssymb,amsmath,nobibnotes,nofootinbib
,superscriptaddress,floatfix]{revtex4}
\usepackage[dvips]{graphicx}
\usepackage{epsfig}
\usepackage{framed}
\usepackage{color}
\usepackage{xcolor}
\usepackage{textcomp}
\usepackage{setspace}
\usepackage{morefloats}
\usepackage[T1]{fontenc}
\usepackage[utf8]{inputenc}
\usepackage{cleveref}
\usepackage{ulem}
\usepackage{comment}
\begin{document}
\title{Flocking and reorientation transition in the 4-state active Potts model}

\author{Swarnajit Chatterjee}
\email{sspsc5@iacs.res.in}
\affiliation{School of Mathematical \& Computational Sciences, Indian Association for the Cultivation of Science, Kolkata -- 700032, India.}

\author{Matthieu Mangeat}
\email{mangeat@lusi.uni-sb.de}
\affiliation{Center for Biophysics \& Department for Theoretical Physics, Saarland University, D-66123 Saarbr{\"u}cken, Germany.}

\author{Raja Paul}
\email{raja.paul@iacs.res.in}
\affiliation{School of Mathematical \& Computational Sciences, Indian Association for the Cultivation of Science, Kolkata -- 700032, India.}

\author{Heiko Rieger}
\email{h.rieger@mx.uni-saarland.de}
\affiliation{Center for Biophysics \& Department for Theoretical Physics, Saarland University, D-66123 Saarbr{\"u}cken, Germany.}



\begin{abstract}
We study the active 4-state Potts model (APM) on the square lattice in which active particles have four internal states corresponding to the four directions of motion. A local alignment rule inspired by the ferromagnetic 4-state Potts model and self-propulsion via biased diffusion according to the internal particle states leads to flocking at high densities and low noise. We compute the phase diagram of the APM and explore the flocking dynamics in the region, in which the high-density (liquid) phase coexists with the low-density (gas) phase and forms a fluctuating band of coherently moving particles. As a function of the particle self-propulsion velocity, a novel reorientation transition of the phase-separated profiles from transversal to longitudinal band motion is revealed, which is absent in the Vicsek model and the active Ising model. We further construct a coarse-grained hydrodynamic description of the model which validates the results for the microscopic model.
\end{abstract}

\maketitle


{\bfseries \itshape Introduction-}
Active matter systems, natural or artificial, consist of a large number of active particles that consume energy to self-propel or exert mechanical forces on the surroundings. While the consumption of energy from the surroundings happens at a single particle level, the interaction between individual particles either occur directly or are mediated by a surrounding medium. The particles exhibit collective motion leading to the spontaneous emergence of synchronized motion of large clusters of self-propelled individuals denoted as flocks, where the typical size of the clusters is significantly larger than the size of an individual \cite{ramaswamy,shaebani,magistris,menon}. 

The flocking transition is an out of equilibrium phenomenon and abundant in nature\cite{marchetti2013}: from human crowds\cite{bottinelli2016,helbing1995}, mammalian herds\cite{garcimartin2015}, bird flocks\cite{ballerini2008}, fish schools\cite{beccoa2006,calovi2014} to unicellular organisms such as amoebae, bacteria\cite{steager2008,peruani2012}, collective cell migration in dense tissues\cite{giavazzi2018}, and sub-cellular structures including cytoskeletal filaments and molecular motors\cite{schaller2010,sumino2012,sanchez2012}, they all show a remarkable active self-organization.
The physics of flocking is also prevalent in nonliving substances such as rods on a horizontal surface agitated vertically\cite{deseigne}, self-propelled liquid droplets\cite{shashi2011}, liquid crystal hydrodynamics, and rolling colloids\cite{bricard2013}.

A widely studied computational model capturing the essential physics of the flocking phenomenon in an active matter system is due to Vicsek \cite{Viscek}. It consists of many self-propelled particles that tend to align with the average direction of the particles in the neighborhood and displays a flocking transition from a strong noise, low-density homogeneous phase to a weak noise, high-density ordered phase. After the landmark emergence of Vicsek Model (VM), Toner and Tu \cite{toner} revealed the nature of the order emerging through flocking via a continuum theory and found that the coherent motion of a flock is a spontaneous broken-symmetry phase without a preferred motion direction. The rich physics of the VM \cite{ginelli} motivated numerous analytical and computational studies that contributed significantly to the understanding of the principles of the flocking transition. They form a wide class of {\it Vicsek-like} models, belonging to the active XY universality class, for polar particles aligning with ferromagnetic interactions\cite{GC2004, TT2005, marchetti, ihle, solon2015, liebchen2017, sandor2017, escaff2018, miguel2018}. Other kinds of particles have also been studied in the literature\cite{peshkov,julicher2018} aligning with nematic interactions such as the self-propelled rods model \cite{bertin2006, bertin2009, marchetti2008, ginelli2010} for polar particles and active nematics \cite{chate2006, bertin2013, bertin2014} for nematic particles.

Recently, in the context of the active Ising model (AIM), Solon and collaborators \cite{ST2013,ST2015,ST2015-2} argued that the flocking transition can be seen as a liquid-gas phase transition rather than an order-disorder transition, similar to the VM \cite{solon2015}. The flocking transition reveals itself as a transition from a disordered gas to a polar ordered liquid through a liquid-gas coexisting phase where polar ordered bands travel in a disordered background. The continuous symmetry of the VM is replaced in the AIM by the discrete symmetry which allows understanding the flocking transition in a simpler and more tractable manner. 

The 4-state active Potts model (APM) addressed in this letter has four internal states corresponding to four motion directions and is defined on a two-dimensional lattice with coordination number $4$. Its two main ingredients leading to flocking are the local alignment interactions and self-propulsion via biased hopping to neighboring sites without repulsive interactions. A 4-state model with hard-core interactions and different transition rules was considered in \cite{peruani2011}, displaying a rich but very different scenario of self-organized patterns. In this letter, we ask how much of the features of the flocking transition of the AIM can also be observed in the APM when particles can move in more than two directions. Consequently, besides a transverse band movement observed in the AIM, are longitudinal band velocities possible in the APM? Is the nature of the transition in the 4-state APM in the universality class of the equilibrium 4-state Potts model? We address these issues by analyzing the flocking transition of the 4-state APM on a square lattice via Monte Carlo simulations, followed by a hydrodynamic theory supporting the numerical results.


{\bfseries \itshape The model-} 
We consider an ensemble of $N$ particles defined on a two-dimensional lattice with coordination number $q$ (square lattice for $q=4$ and triangular lattice for $q=6$). Each particle is in one of $q$ discrete internal states corresponding to a movement in one of the $q$ lattice directions. Each site can be occupied by multiple particles and each particle can either flip to a different spin state or hop to the nearest neighbor site. The alignment ({\it i. e.} flip) probability for particles on site $i$ are defined by the local Hamiltonian
\begin{equation}
\label{Hpotts}
H_i= - \frac{J}{2\rho_i}\sum_{k=1}^{\rho_i}\sum_{l \neq k}(q\delta_{\sigma_k,\sigma_l}-1),
\end{equation}
where the double sum runs over all particle pairs $(k,l)$ in site $i$, $\sigma_k$ denotes the state of the $k^{\rm th}$ particle on site $i$: $\sigma_k \in \{1,\ldots,q\}$, and $J$ is the strength of the coupling between different particles on the same site. The number of particles on site $i$ is $\rho_i=\sum_{\sigma=1}^{q}n_i^{\sigma}$ with $n_i^\sigma=\sum_{k=1}^{\rho_i} \delta_{\sigma_k,\sigma}$, the number of particles in state $\sigma$. The local magnetization in the direction ${\sigma}$ on site $i$ is 
\begin{equation}
\label{defmag}
m_i^{\sigma}=\sum_{k=1}^{\rho_i}\frac{(q\delta_{\sigma,\sigma_k}-1)}{q-1}=\frac{q n_i^\sigma - \rho_i}{q-1}.
\end{equation}
For $q=2$, we retrieve the expression of local Hamiltonian and magnetization for the AIM \cite{ST2015}. A particle on site $i$ in state $\sigma$ changes its state to $\sigma'$ with the rate $W_{\rm flip}(\sigma, \sigma') \propto \exp(-\beta \Delta H_i)$, where $\Delta H_i={qJ}(n_i^\sigma-n_i^{\sigma'}-1)/{\rho_i}$. Moreover, the particle performs a biased diffusion and jumps to the neighboring site in the direction $p\in\{1,...,q\}$ with the rate $W_{\rm hop}(\sigma,p) = D[1+\epsilon(q\delta_{\sigma,p}-1)/(q-1)]$ where $\epsilon$ is the self-propulsion parameter. 

A Monte Carlo (MC) simulation of the stochastic process defined in this way evolves in unit Monte Carlo steps (MCS) $\Delta t$ resulting from a microscopic time $\Delta t/N$, $N$ being the total number of particles. During $\Delta t/N$, a randomly chosen particle either update its spin state with probability $p_{\rm flip} = \sum_{\sigma' \ne \sigma} W_{\rm flip}(\sigma, \sigma') \Delta t$ or hop to one of the $q$ neighboring sites with probability $p_{\rm hop} = \sum_{p=1}^q W_{\rm hop}(\sigma,p) \Delta t= q D \Delta t$. The probability that nothing happens during this time is represented by $p_{\rm wait}=1-p_{\rm flip} - p_{\rm hop}$. An expression for $\Delta t$ can be obtained by minimizing $p_{\rm wait}$: $\Delta t = [qD+\exp(q\beta J)]^{-1}$.


{\bfseries \itshape Results-} 
The 4-state APM is studied on a square lattice of linear dimensions $L=200$ with periodic boundary conditions, where individual particle states $\sigma=\{1,2,3,4\}$ correspond to the movement directions right, up, left and down, respectively. In the following, we set the diffusion constant $D=1$ and the coupling constant $J=1$. Remaining parameters $\beta = 1/T$ regulates the strength of the noise, $\rho_0=N/L^2$ defines the average particle density (i.e. average number of particles at a given site), and $\epsilon$ controls the self-propulsion velocity of the particles.

\begin{figure}[t]
\centering
\includegraphics[width=\columnwidth]{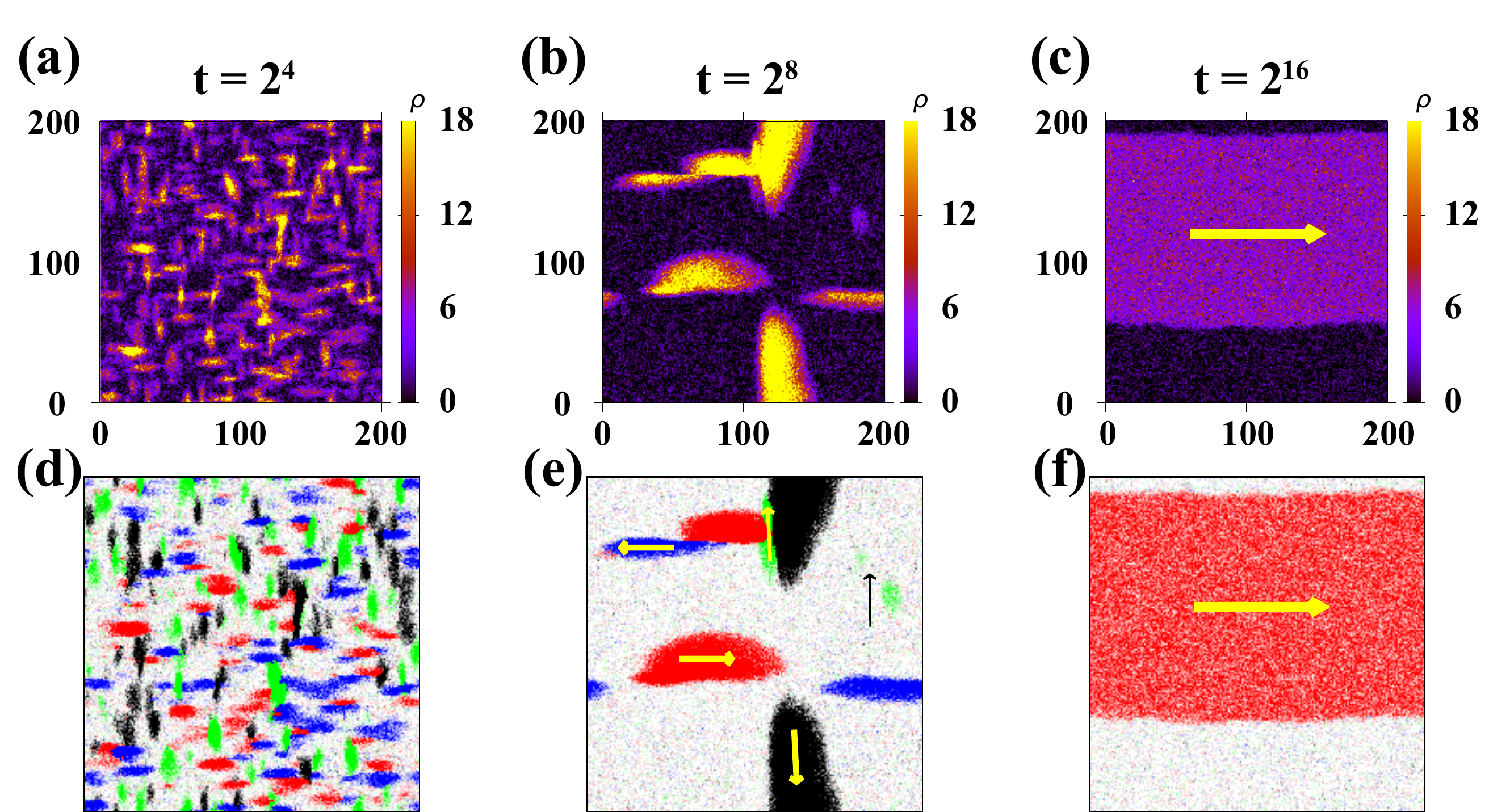}
\caption{(Color online) Flocking process in the 4-state APM: Snapshots of the evolution for $\epsilon=2.2$, $\beta=0.8$ and $\rho_0=4$ at times $t=2^4$ MCS, $t=2^8$ MCS, and $t=2^{16}$ MCS (from left to right). {\bf(a--c)} Particle density, color coded according to the color bar on the right. {\bf(d--f)} Particles move in 4 directions corresponding to the snapshots (a--c). Ref: red: right, green: up, blue: left, black: down, white: empty.}
\label{fig1}
\end{figure} 

In Fig.~\ref{fig1}, snapshots demonstrate the evolution of the flocking in the 4-state APM starting from a random initial configuration. One sees that, first, many clusters of particles move in all $4$ directions. Upon collision, these clusters coalesce and grow in size, until a single band is formed, in which all particles move in the same direction. Interestingly, a spontaneous longitudinal motion of the band is observed, in contrast to the transverse motion of the AIM \cite{ST2013,ST2015,ST2015-2}, which will be analyzed further below.

\begin{figure}[t]
\centering
\includegraphics[width=\columnwidth]{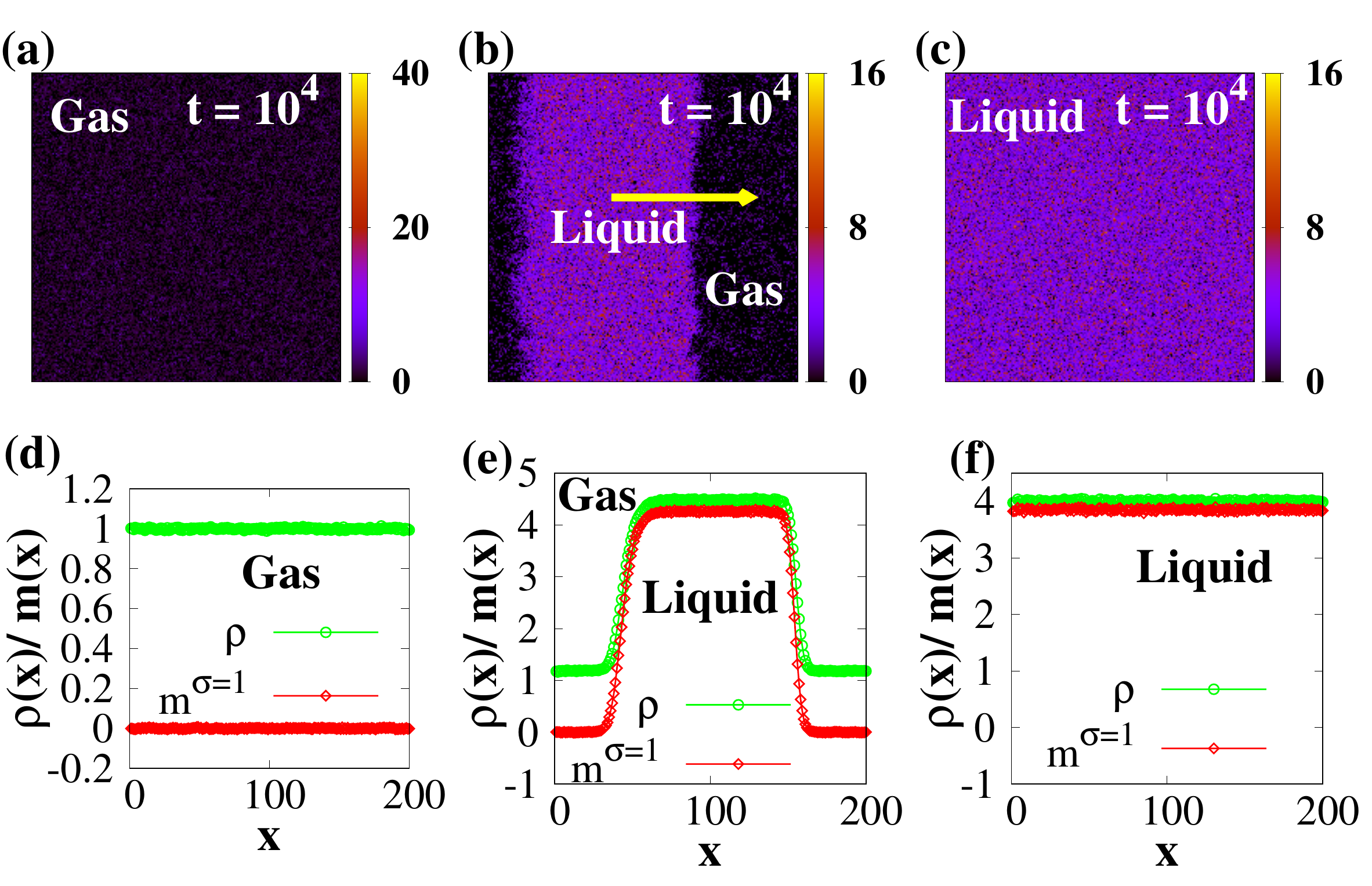}
\caption{Three different phases of APM for $\epsilon = 0.9$. Instantaneous snapshots at $t=10^4$ MCS display gas, coexisting liquid-gas and liquid phases in {\bf (a)}, {\bf (b)}, and {\bf (c)} respectively. The disordered gaseous state of {\bf (d)} at high temperature and low-density ($\beta = 0.7$, $\rho_0 = 1$), passes through a stable liquid-gas coexistence phase {\bf (e)} at intermediate temperature and density ($\rho_0 = 3$ and $\beta = 0.8$) and becomes a polar ordered liquid phase {\bf (f)} at low temperature and high-density ($\beta = 0.9$, $\rho_0 = 4$).}
\label{fig2}
\end{figure}

Fig.~\ref{fig2} shows snapshots from the stationary states and characterizes the three different phases. The snapshots shown in (a), (b), and (c) correspond to the magnetization and density profiles shown in (d), (e), and (f) with $\epsilon=0.9$. A disordered gaseous phase appears in (d) at a relatively high temperature ($\beta=0.7$) and low-density ($\rho_0=1$) with average local magnetization $\langle m_i^{\sigma=1} \rangle \sim 0$. Characteristics of a polar ordered liquid phase (f), observed at a relatively low temperature ($\beta=0.9$) and high-density ($\rho_0=4$), where the equilibrated average magnetization ($\langle m_i^{\sigma=1} \rangle \neq 0$) nearly equals the average density. The transition from a gas to a polar liquid phase occurs through a liquid-gas coexistence phase (e) at intermediate density ($\rho_0=3$) and temperature ($\beta=0.8$), where the density and magnetization of the liquid phase can be significantly large depending on the average particle density $\rho_0$. In this phase, the band of polar liquid propagates transversely on a disordered gaseous background. 

\begin{figure}[t]
\centering
\includegraphics[width=\columnwidth]{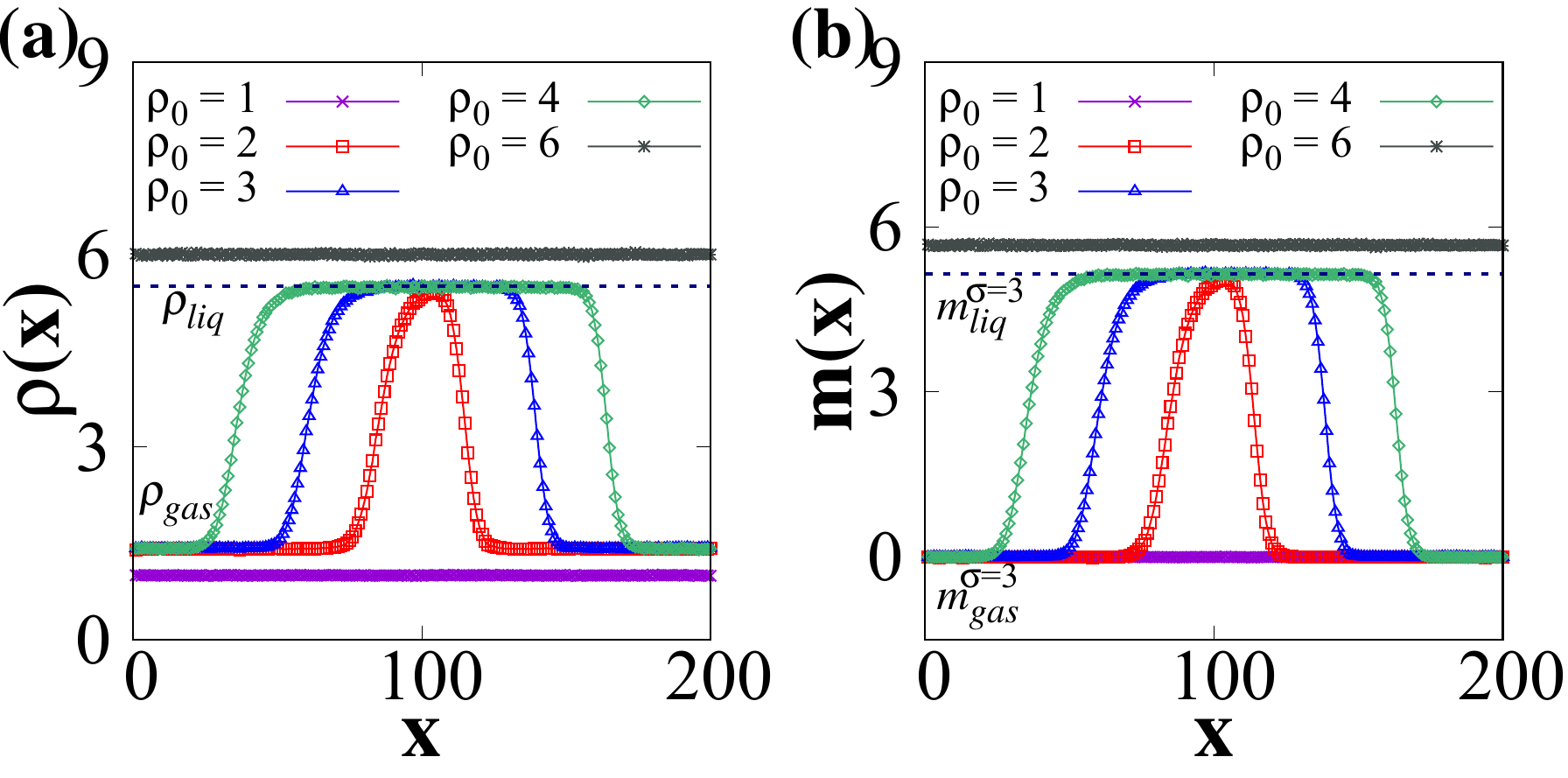}
\caption{Phase-separated average density {\bf (a)} and $\sigma=3$ state magnetization {\bf (b)} profiles of liquid ($\rho_{\rm liq}$) and gaseous ($\rho_{\rm gas}$) phases with increasing initial $\rho_0$ are shown. A complete gaseous and liquid phases correspond to $\rho_0$ = 1 and 6 respectively, keeping $\epsilon = 0.9$ and $\beta = 0.7$ fixed.}
\label{fig3}
\end{figure}

Segregated density and magnetization profiles of the liquid-gas coexistence phase in Fig.~\ref{fig3}(a) and Fig.~\ref{fig3}(b) respectively, suggest that the width of the polar liquid band increases with the initial average density $\rho_0$ and behaves as in the AIM \cite{ST2013,ST2015,ST2015-2}. Interestingly, the broadening of the band does not affect the equilibrium densities of the liquid ($\rho_{\rm liq}$) and the gaseous ($\rho_{\rm gas}$) phases. Notice, the non-vanishing magnetization of the liquid band ($m_{\rm liq} \neq 0$) in Fig.~\ref{fig3}(b) approaches the density of the liquid phase indicating the dominance of a single internal state within the band. In low-density gaseous phase, magnetization vanishes completely ($m_{\rm gas} = 0$) due to the mixing of all states with equal probability. Fig.~\ref{fig3} also demonstrates the phase transition from the gas to the co-existence phase to the liquid phase for constant thermal noise and increasing $\rho_0$. Keeping the parameter space constant, we get the gaseous phase for $\rho_0=1$ whereas $\rho_0=6$ gives rise to the liquid phase.

\begin{figure}[t]
\centering
\includegraphics[width=\columnwidth]{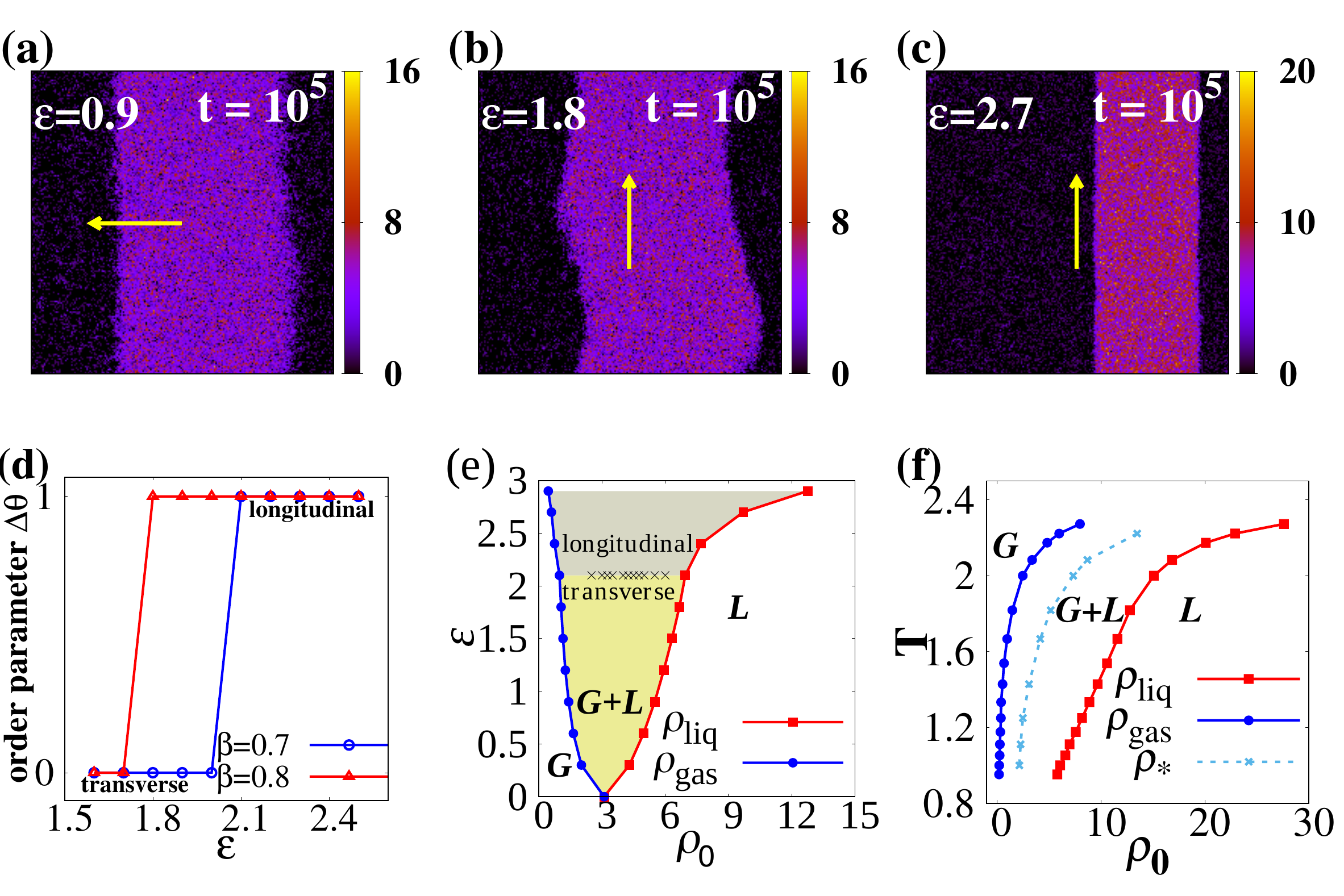}
\caption{{\bf (a--c)}~Snapshots of the stationary bands for $\beta=0.8$ and $\rho_0=3$ at $t=10^5$ MCS for {\bf (a)}~$\epsilon=0.9$, {\bf (b)}~$\epsilon=1.8$, and {\bf (c)}~$\epsilon= 2.7$. Colorbar represents the local particle-density and arrows indicate the direction of motion of the particles. For $\epsilon=0.9$, the particle moves transversal to the band orientation, whereas for $\epsilon=1.8$ and 2.7, the particle movement direction is longitudinal. {\bf (d)} Orientational order parameter $\Delta\theta$, measuring the average difference between the band orientation and particle movement direction, as a function of the effective velocity $\epsilon$ for $\beta = 0.7$ and $0.8$, and $\rho_0=3$. {\bf (e--f)} Phase diagrams of the APM. {\bf (e)} Phase diagram in the $(\epsilon,\rho_0)$ plane for $\beta=0.7$. The reorientation transition line from transverse to longitudinal particle motion in the stationary bands, indicated by a black dotted line, is independent of $\rho_0$. {\bf (f)} In the $(T,\rho_0)$ plane the solid curves with circles ($\rho_{\rm gas}$) and squares ($\rho_{\rm liq}$) at $\epsilon$ = 2.7, separate three phases: gas (G), gas-liquid coexistence (G+L) and liquid (L) as $\rho_0$ increases. The dotted line indicates the transition density $\rho_*$ for $\epsilon=0$.}
\label{fig4}
\end{figure}

A novel feature of the flocking phenomenon in the APM is documented in Fig.~\ref{fig4}: The snapshots in (a--c) show the orientation of the band formed in the stationary state for increasing values of the effective velocity. For $\epsilon = 0.9$, the particles in the band move in the transverse direction (implying that the whole band moves), whereas for $\epsilon=1.8$ and $2.7$ the particles move in the longitudinal direction (implying that the band is immobile up to fluctuations).

To ensure that our findings are not artifacts of any algorithmic implementation, we studied the $q=6$ of the $q$-state APM \cite{tbp}. The 6-state APM lives on a triangular lattice with periodic boundary conditions to make the number of states and biased hopping directions equivalent, without altering the expressions of $W_{\rm flip}$ and $W_{\rm hop}$. In complete analogy to the $q=4$ scenario, we observe a reorientation transition from transversal motion of co-existence band at small $\epsilon$ to longitudinal motion at large $\epsilon$ for a fixed $\beta$. An extensive study of the 6-state APM will be presented elsewhere \cite{tbp}. 

We further emphasize that in the simulations of the 4-state APM, irrespective of the initial conditions (e.g., completely random or a vertical band or a horizontal band or a square bubble of liquid on a gaseous background), for large $\epsilon$ (greater than a threshold $\epsilon=\epsilon_r$, representing the reorientation transition), we always get a longitudinal motion of the self-propelled band. To quantify this transition we introduce the orientational order parameter, $\Delta\theta = \langle\cos(\theta-\phi_k)\rangle$, where the angle $\theta$ measures the band orientation, $\phi_k$ the movement direction of individual particles $k$ inside the band and $\langle\ldots\rangle$ is the average over time and particles. The data shown in Fig.~\ref{fig4}(d) as a function of the effective velocity $\epsilon$ for fixed $\beta$ and $\rho_0$ demonstrate a discontinuous transition that occurs at larger values for $\epsilon$ with increasing $\beta$. The transition of the band from the transverse to the longitudinal orientation is caused by the large difference of the particle hopping rates due to $\epsilon$. For instance, at $\epsilon=2.7$, a particle in the state $\sigma = 1$ hops to the right with rate 3.7, and with rate 0.1 in the other three directions. Therefore a vertical band consisting of the particle in state $\sigma = 1$, will move faster in the ``right" direction causing the vertical band to collapse and the subsequent reorientation along the horizontal direction. 

Based on the orientational order parameter we determined the threshold values of $\epsilon_r$ where the reorientation transition happens and inserted them into the $(\epsilon,\rho_0)$-phase diagram shown in Fig.~\ref{fig4}(e). At a fixed $\rho_0$, the binodals $\rho_{\rm gas}$ and $\rho_{\rm liq}$ are computed from the time-averaged phase-separated density profiles. It turns out to be a horizontal line for fixed $\beta$, i.e. independent of $\rho_0$. Two binodals separate the three phases, gas, liquid and gas-liquid coexistence region, and merges at the density $\rho_*$ (e.g., $\rho_*\simeq 3.10$ when $\epsilon=0$ at $\beta=0.7$). For completeness, Fig.~\ref{fig4}(f) shows the temperature-density $(T,\rho_0)$ phase diagram for $\epsilon=2.7$ where $\rho_{\rm gas}$ and $\rho_{\rm liq}$. The dashed line represents $\rho_*$ at $\epsilon=0$ which manifests a direct order-disorder phase transition without going through a co-existence regime.


Next, we checked whether the flocking and reorientation transition 
observed in the APM can also be predicted by a hydrodynamic continuum theory. From the microscopic hopping and flipping rules, we obtain the Master equation
\begin{align}
\partial_t \langle n_i^\sigma \rangle &= D \left(1 - \frac{\epsilon}{3} \right) \sum_{p=1}^4 \left[ \langle n_{i-p}^\sigma \rangle - \langle n_i^\sigma \rangle \right] \nonumber \\
&+ \frac{4 D \epsilon}{3} \left[ \langle n_{i-\sigma}^\sigma \rangle - \langle n_i^\sigma \rangle \right] \nonumber\\
&+ \sum_{\sigma' \ne \sigma } \left\langle n_i^{\sigma'} W_{\rm flip}(\sigma',\sigma) - n_i^\sigma W_{\rm flip}(\sigma,\sigma') \right\rangle
\end{align}
For small lattice spacing $a \simeq 1/L$, the hydrodynamic equation in the large system size limit $L\to\infty$ can be derived for the average density of particle in the state $\sigma$: $\rho_\sigma({\bf x},t) = \langle n_i^\sigma(t) / L^2 \rangle$, where the position ${\bf x}$ corresponds to the site $i$ \cite{tbp}. We only keep the first-order terms in the $m_\sigma \ll \rho$ expansion, where $m_\sigma = (4\rho_\sigma-\rho)/3$ is the magnetization of the state $\sigma$ from Eq.~(\ref{defmag}). Assuming all magnetizations $m_\sigma$ are identically distributed Gaussian variables with variance $\alpha_m \rho$ proportional to the local mean population (which we verified by MC simulations of the microscopic model), we obtain up to order ${\cal O}(m_\sigma^3)$ \cite{tbp}
\begin{gather}
\label{eqAPMHydro}
\partial_t \rho_\sigma = D_\parallel \partial_\parallel^2 \rho_\sigma + D_\perp \partial_\perp^2 \rho_\sigma - v \partial_\parallel \rho_\sigma + \\
\sum_{\sigma' \ne \sigma } \left[\frac{4\beta J}{\rho}(\rho_\sigma+\rho_{\sigma'}) -1 - \frac{r}{\rho} - \alpha \frac{(\rho_\sigma-\rho_{\sigma'})^2}{\rho^2}\right](\rho_\sigma-\rho_{\sigma'}),\nonumber
\end{gather}
where $\alpha= (4 \beta J)^2(1-2\beta J/3)/2$ depends on the temperature and $r= 27\alpha_m \alpha/8$ is a new positive parameter. $D_\parallel$ and $D_\perp$ are the diffusion constants in the parallel direction ${\bf e_\parallel} = (\cos \phi, \sin \phi)$ and orthogonal direction ${\bf e_\perp} = (\sin \phi, -\cos \phi)$, respectively, with $\phi = (\sigma-1) \pi/2$ the angle of the direction of motion in state $\sigma$. Correspondingly $\partial_\parallel$ and $\partial_\perp$ are the derivatives in the parallel and orthogonal direction, respectively, $\partial_\parallel= {\bf e_\parallel} \cdot \nabla$ and $\partial_\perp= {\bf e_\perp} \cdot \nabla$. Note that the simple mean-field theory, neglecting fluctuations (i.e. setting $r=0$), does not predict stable phase-separated profiles and only gives the trivial homogeneous solution, as in the AIM \cite{ST2013,ST2015}. In the following, we analyze the refined mean-field theory (\ref{eqAPMHydro}) and set $D=1$, $J=1$ and $r=1$ defining a scaling for the time, temperature and density.

\begin{figure}[t]
\centering
\includegraphics[width=\columnwidth]{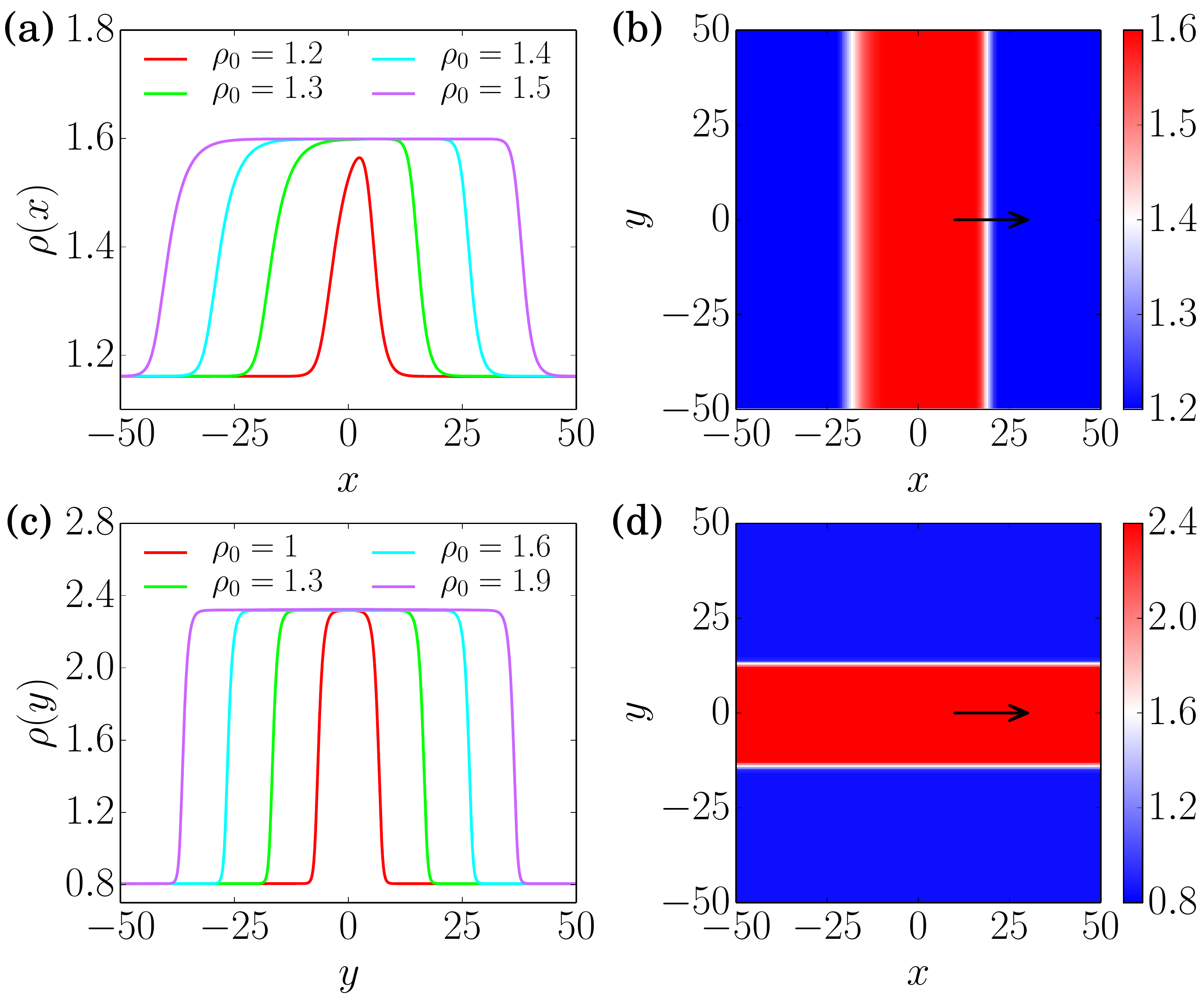}
\caption{Reorientation transition from hydrodynamics. {\bf (a--b)} Phase-separated density profiles for $\epsilon=0.3$ and $\beta=0.75$ exhibiting a transverse band motion where the gas density is $\rho_{\rm gas}=1.16$ and the liquid density is $\rho_{\rm liq}=1.60$. {\bf (c--d)} Phase-separated density profiles for $\epsilon=2.5$ and $\beta=0.75$ exhibiting a longitudinal band motion with $\rho_{\rm gas}=0.805$ and $\rho_{\rm liq}=2.32$. {\bf (a)} and {\bf (c)} represent the marginal density $\rho(x)$ and $\rho(y)$ respectively for several values of $\rho_0$. {\bf (b)} and {\bf (d)} demonstrate the reorientation transition for $\rho_0=1.33$ as we increase the value of $\epsilon$. Red denotes liquid phase and blue denotes gas phase.} 
\label{fig5}
\end{figure}

We solve Eq.~(\ref{eqAPMHydro}) numerically using FreeFem++\cite{ff}, a software based on the finite element method\cite{fem}, for discrete-time and on a regular triangular mesh-grid. We obtain the disordered gas, polar ordered liquid, and liquid-gas coexisting phase as for the microscopic model, together with transversely and longitudinally moving bands (data not presented here). The reorientation transition observed above in the microscopic model is also predicted by our hydrodynamic theory, as is shown in Fig.~\ref{fig5} for $\beta=0.75$. Several phase-separated density profiles averaged along the $y$-axis and along the $x$-axis are shown in Fig.~\ref{fig5}(a) and Fig.~\ref{fig5}(c), respectively, for increasing values of $\rho_0$. For a density $\rho_0=1.33$, Fig.~\ref{fig5}(b) shows the transversal band motion for $\epsilon=0.3$, where band orientation and band propulsion are mutually orthogonal, whereas the longitudinal motion is seen in Fig.~\ref{fig5}(d) for $\epsilon=2.5$. Note that the colorbars in Fig.~\ref{fig5}(b) and Fig.~\ref{fig5}(d) represent particle density where red denotes high-density liquid and blue denotes low-density gas. The origin of the reorientation transition lies in the decrease of the orthogonal diffusivity $D_\perp$ for increasing self-propulsion velocity $\epsilon$ that vanishes at the maximum value of $\epsilon=3$, leading to the longitudinal band motion for large $\epsilon$ (corresponding stability analysis will be presented in \cite{tbp}). At small velocities, parallel and orthogonal diffusivities do not differ much and transverse band motion emerges as in the AIM and the VM.

\begin{figure}[t]
\centering
\includegraphics[width=\columnwidth]{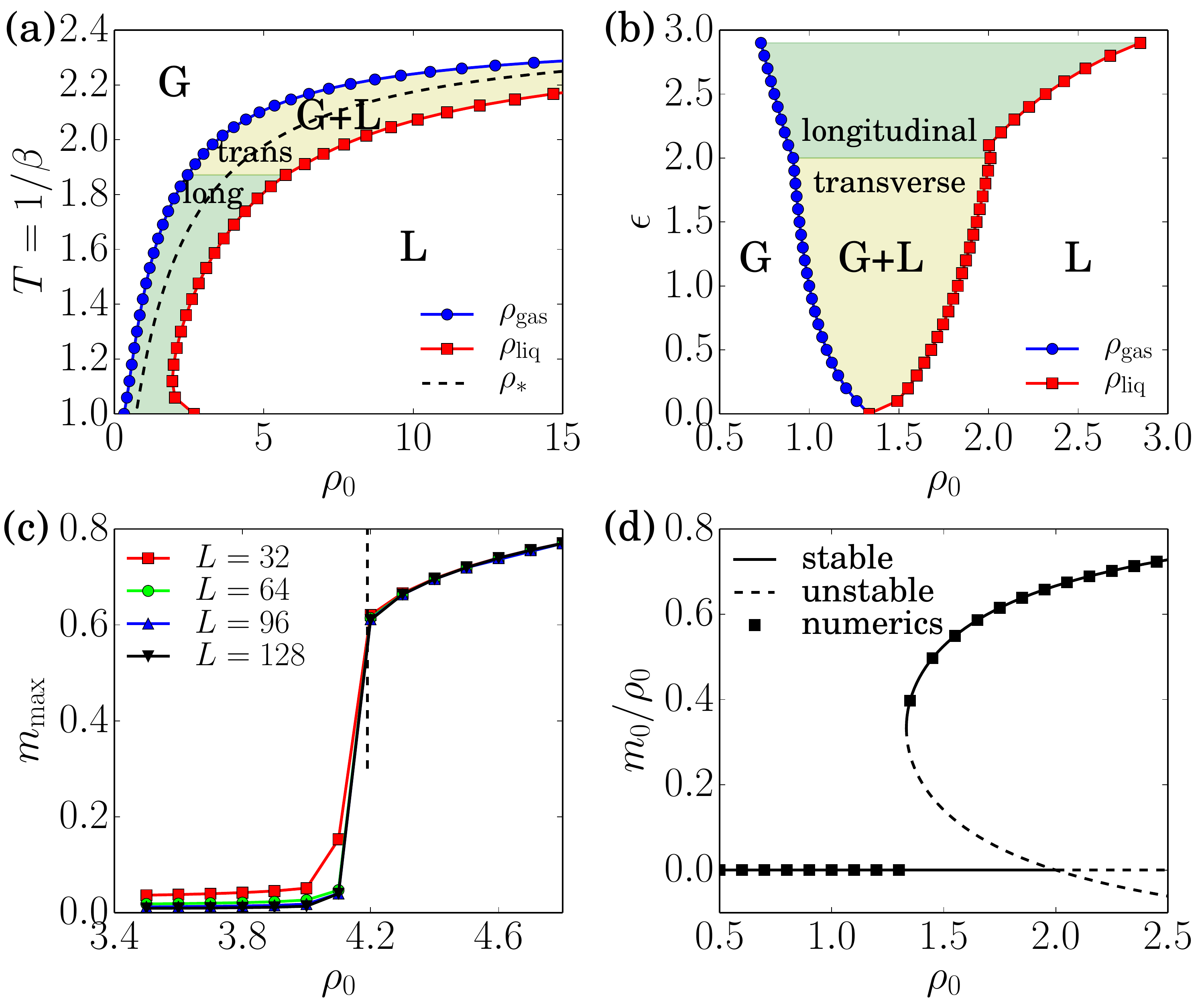}
\caption{{\bf(a--b)} Phase diagrams for the hydrodynamic equations of the APM. The solid lines represent the coexistence densities $\rho_{\rm gas}$ and $\rho_{\rm liq}$ of the gas and liquid phases. {\bf (a)}~Temperature-density phase diagram for the physical parameters $\epsilon=2.5$ and $L=100$. Reorientation transition happens at $T=1.9$ from longitudinal motion at low $T$ to a transverse motion at high $T$. The dotted line represents the ordered-disordered transition density $\rho_*$ at $\epsilon=0$ and the critical point is reached for $T_c \simeq 2.417$ and $\rho_c=+\infty$. {\bf (b)}~Velocity-density phase diagram for $\beta=0.75$ and $L=100$ where the reorientation transition happens at $\epsilon=2.0$. A first-order phase transition takes place at $\epsilon=0$ between ordered and disordered phases at density $\rho_*=4/3$. {\bf (c--d)} Characterization of disordered-ordered phase transition in the purely diffusive 4-state APM. {\bf (c)}~Normalized maximal magnetization $m_{\rm max}$ versus $\rho_0$ for lattice size $L$ = 32, 64, 96, and 128 is shown for $\beta= 0.6$ and $\epsilon=0$. The sudden jump in the magnetization signals a possible first-order phase transition. {\bf (d)}~Steady-state magnetization $m_0/\rho_0$ versus $\rho_0$ for $\beta =0.75$ and $\epsilon=0$. A first-order phase transition takes place when the ordered phase appears at $\rho_0=\rho_*=4/3$. The numerical values of $m_0$ obtained with FreeFem++ are shown with squares for an inhomogeneous initial condition.}
\label{fig6}
\end{figure}

In Fig.~\ref{fig6}(a) and Fig.~\ref{fig6}(b), the temperature-density and velocity-density phase diagrams are represented respectively for $\epsilon=2.5$ and $\beta = 0.75$ with binodals and corresponding band motions obtained using FreeFem++\cite{ff}. The gas and liquid densities depend on the temperature $T=1/\beta$ and the self-propulsion velocity $\epsilon$ and a phase-separated profile can exist for $\epsilon>0$ and $\beta > 1- \sqrt{22}/8 \simeq 0.413$. Note that the limit $T_c \simeq 2.417$ is close to the one observed for the microscopic system shown in Fig.~\ref{fig4}(f) and is analogous to a liquid-gas critical point at infinite density ($\rho_c=+\infty$) since the disordered phase could not transform continuously into the ordered phase due to the $Z_4$ symmetry breaking as for AIM\cite{ST2015} and VM\cite{solon2015}, explaining the absence of the supercritical region. This behavior is given by the ordered-disordered transition line $\rho_*$ at $\epsilon=0$, defined by the existence of ordered solutions of Eq.~(\ref{eqAPMHydro}). Two different inhomogeneous profiles are stable for the APM: a transverse band of polar liquid at small $\epsilon$ and large $T$ and a longitudinal of polar liquid at large $\epsilon$ and small $T$. The values of the binodals are not continuous at the reorientation transition which takes place for $T=1.33$ ($\beta=0.75$) at $\epsilon=2.0$ on Fig.~\ref{fig6}(b) and for $\epsilon=2.5$ at $T=1.9$ on Fig.~\ref{fig6}(a).

The data presented in Fig.~\ref{fig6}(c) shows the normalized maximal magnetization among the four internal states, denoted $m_{\rm max}$, against density $\rho_0$ for $\beta=0.6$ and $\epsilon=0$, where a jump in the magnetization occurs around the transition point at $\rho_0=\rho_*$. Among the four different magnetizations corresponding to four internal states, we consider the maximum: $m_{\rm max} = \max_\sigma \langle m_i^\sigma/ \rho_i \rangle$ and plotted against $\rho_0$. This discontinuity becomes sharper with increasing system sizes and this discontinuous change of a large $m_{\rm max}$ at a high-density ordered phase to a small $m_{\rm max} \simeq 0$ at a relatively lower density indicates the possibility of a first-order transition. Ideally, a fully ordered state should acquire magnetization $m_{\rm max} \simeq 1$, however, $m_{\rm max}$ ($\rho_0=4.8) \simeq 0.77$ in the ordered liquid phase suggests that all the particles on a lattice site may not belong to the same $\sigma$ and therefore one realizes that from Eq.~(\ref{defmag}), $m_{\rm max} <1$. 

An important observation here is that the $\epsilon=0$ order-disorder transition of the APM does not appear to be in the same universality class as the (passive) 4-state Potts model, whereas the $\epsilon=0$ critical point of the AIM was found to be in the Ising universality class \cite{ST2015}. Both models, AIM and APM differ from their conventional (passive) counterparts in so far as only particles on the same site interact ferromagnetically in the AIM and APM, whereas in the passive Ising and Potts model spins on neighboring sites interact ferromagnetically. For the conventional (passive) $q$-state Potts model, the temperature-driven transition is continuous for $q \leqslant q_c$ and first-order for $q > q_c$, with $q_c=4$ for the square-lattice with nearest-neighbor interactions and $q_c \simeq 2.8$ for the simple-cubic lattice \cite{baxter,wu,hartmann}. Fig.~\ref{fig6}(d) shows a discontinuous jump in the magnetization between the ordered and the disordered phases at $\epsilon=0$, characteristic of a first-order phase transition. From Eq.~(\ref{eqAPMHydro}), two ordered solutions can be extracted when $\rho_0>\rho_*$: 
\begin{equation}
\frac{m_0^\pm}{\rho_0} = \frac{\beta J}{\alpha} \left(1 \pm \sqrt{1+ \frac{(2\beta J -1 - r/\rho_0)\alpha}{(\beta J)^2}}\right),
\end{equation}
and only the solution $m_0^+$ is linearly stable at $\epsilon=0$ \cite{tbp}.


{\bfseries \itshape Discussion-}
In conclusion, we analyzed the flocking transition in the two-dimensional 4-state APM. We found a novel reorientation transition where the orientation of the liquid bands emerging in the gas-liquid coexistence region switches from transversal (with respect to the average particle movement direction) for small effective velocities to longitudinal for large effective velocities. This reorientation transition is a novel feature of the APM which is absent in the AIM \cite{ST2015} and in the VM \cite{solon2015} where bands always move in the transverse direction. It has been shown \cite{solon2015} that in the AIM the inhomogeneous polar liquid bands are fully phase-separated, whereas in the VM a smectic {\it microphase} separation emerges in the coexistence region, which means that the width of the polar band has a maximal size leading to a succession of several bands. The transversal orientation of the bands emerging in the VM has been understood within a hydrodynamic theory \cite{bertin2009}, which predicts that the long-wavelength instability is stronger in the broken symmetry direction. On the other hand, in a system with self-propelled rods and in active nematics the long-wavelength instability is stronger in the perpendicular direction with respect to the collective motion and this eventually gives rise to longitudinal bands \cite{bertin2013,bertin2014,marchetti2008,ginelli2010}. There is also a theoretical understanding of the relation between band orientation and broken symmetry direction of the three main classes of active matter, namely the Vicsek-like models, active nematics, and self-propelled rods \cite{peshkov}.

In contrast to the VM and AIM and active nematics, the APM exhibits both types of collective motion: transverse and longitudinal motion of the ordered band. It should be emphasized, that although active Brownian particles with repulsive interactions and Vicsek-like alignment rules \cite{peruani2011, farrell2012, MG2018} also display both longitudinal and transverse band motions, for {\it small} P\'eclet numbers (small velocities) and {\it large} P\'eclet numbers, respectively, exactly opposite to what happens in the APM, see Fig.~\ref{fig4}.

The reorientation transition is discontinuous and does not depend on the average particle density. Moreover, the zero-velocity transition of the APM is a first-order transition and thus does not fall in the same universality class as the passive 4-state Potts model, in contrast to the $q=2$ case, the AIM. A hydrodynamic continuum theory describes the flocking transition of the microscopic model and predicts the reorientation transition as a consequence of the existence of two different diffusion constants: one for the direction parallel to the motion of the particle and one perpendicular. Preliminary results for the 6-state APM on the triangular lattice bears the signature of $q=4$, including the reorientation transition (details will be reported in \cite{tbp}, where the limit $q \to \infty$ will also be explored). Furthermore, a generalization to repulsive interactions among particles on the same site bears the potential for an even larger variety of self-organization patterns.


\begin{acknowledgments}
S.C. is thankful to Alexandre P. Solon for helpful discussions and CSIR, India (through Grant No. 09/080(0897)/2013-EMR-I) and Indian Association for the Cultivation of Science for financial support. R.P. thanks CSIR, India, for support through Grant No. 03(1414)/17/EMR-II and the 
SFB 1027 for supporting his visit to the Saarland University for discussion and finalizing the project. M.M. and H.R. were financially supported by
German Research Foundation (DFG) within the Collaborative Research Center SFB 1027.
\end{acknowledgments}

\def\bibsection{\section*{\refname}}

\end{document}